\documentclass{article}
\usepackage{graphicx}
\usepackage{epsfig}
\usepackage{amsfonts}
 \usepackage{amssymb}
\usepackage[dvips]{color}
\date{\today}

\newcommand{\insertplot}[5]{\begin{figure}
 \hfill\hbox to 0.05in{\vbox to #5in{\vfill
 \inputplot{#1}{#4}{#5}}\hfill}
 \hfill\vspace{-.1in}
 \caption{#2}\label{#3}
 \end{figure}}
 \newcommand{\inputplot}[3]{
 \special{ps: plotfile #1}
}
\newcounter{fig}

\newcommand{\ee}{\end{equation}}
\newcommand{\eea}{\end{eqnarray}}
\newcommand{\be}{\begin{equation}}
\newcommand{\bea}{\begin{eqnarray}}

\begin{document}

 \title{Born-Infeld stars and charged black holes \\ surrounded by scalar clouds
}

\author{
{\large Yves Brihaye}$^{\dagger}$, and
{\large Betti Hartmann} $^{\ddagger}$
\\
\\
$^{\dagger}${\small Physique de l'Univers, Universit\'e de
Mons, 7000 Mons, Belgium}
\\
$^{\dagger}${\small Department of Mathematics, University College London, Gower Street, London, WC1E 6BT, UK}
}
\maketitle
\begin{abstract}
We discuss the formation of scalar clouds on charged, spherically symmetric and static stars and black holes.
We first discuss Reissner-Nordstr\"om black holes with electric and magnetic charge 
and present new results demonstrating the existence of a second branch of solutions. Moreover, we find that the presence of the magnetic charge allows for smaller mass and Noether charge when the clouds are close to their minimal possible mean radius. Replacing standard electrodynamics
by Born-Infeld (BI) electrodynamics, the background model possesses globally regular, star-like as well as black hole solutions. We show that the former can be scalarized as well and that the scalar clouds become more compact when decreasing the electric charge of the BI star. Finally, we discuss a magnetic dipole field in the background of the scalarized BI star and show that the strong gravitational field of this star leads to 
a significant change. 
  \end{abstract}

\section{Introduction}
Compact objects and their (often) strong gravitational fields are of interest when studying the foundations and possible modifications of General Relativity. When considering static spherically symmetric and asymptotically flat solutions of the combined Einstein-Maxwell equations, a black hole solution exists. This is the Reissner-Nordstr\"om (RN) solution \cite{RN} which is necessarily static \cite{jebsen,israel} and described by only two parameters subject to a Gauss law: the mass $M$ and the total (electric and/or magnetic) charge $Q$ \cite{carter1971,robinson, heusler, MTW}. The electric field of the charge diverges at the location of the charge and similar problems appear for the Dirac magnetic monopole field. However, since these singularities are shielded by an event horizon, one
can construct black hole solutions regular outside the horizon. In \cite{Born:1934gh} an alternative model to Maxwell's electrodynamics was introduced, often called {\it Born-Infeld (BI) electrodynamics}, which 
cures the problem of point-like electric charges. In fact, the electric field is well-behaved at the location of the charge. Consequently, regular black holes have been constructed in Einstein-BI models 
\cite{Ayon-Beato:1998hmi,Ayon-Beato:1999kuh, Ayon-Beato:1999qin, Bronnikov:2000vy, Burinskii:2002pz}
with both electric and magnetic charges (see also \cite{Yang:2022qoc} for a recent construction). Moreover, globally regular solutions do exist \cite{Dymnikova:2004zc} often referred to as ``gravastars'' \cite{Lobo:2006xt}. 

While a number of scalar no-hair theorems for asymptotically flat and stationary black holes exist \cite{Herdeiro:2015waa}, there are possibilities to circumvent these. Static black hole solutions
of the Einstein-Maxwell equations can carry non-trivial scalar hair for a self-interacting complex scalar field with non-trivial time-dependence of the form $\sim\exp(i\omega t)$, where $\omega$ a real constant. For the scalar field to be non-trivial on and outside the event horizon, the synchronization condition
$\omega = qV(r_h)$ has to be fulfilled, where $q$ is the gauge coupling and $V(r_h)$ the electric potential on the horizon.
These solutions were presented in \cite{Hong:2019mcj, Herdeiro:2020xmb} and additional results were given
in \cite{Brihaye:2020vce}. In this paper, we present the domain of existence of solutions in the $q$-$Q$-plane, see Section 3. Recently, the results for electrically charged black holes were extended to include a magnetic charge of the RN solution \cite{Herdeiro:2024yqa} and the backreaction of the scalar field on the space-time was investigated.
In this paper we extend these results by demonstrating that for a fixed space-time background two branches of solutions exist, see Section 3, and explain the difference between the two branches. 
Moreover, in Section 4, we discuss how the structure of solutions is changed when replacing
standard electrodynamics by BI electrodynamics. In particular, we discuss the globally regular, star-like solution in detail and show that this can also be scalarized. Section 5 contains our Conclusions, while the Appendix shows how the strong gravitational field of the BI star influences a magnetic dipole field.

\section{The set-up}
We consider a number $n+1$ of self-interacting, massive and complex scalar fields $\Phi_k$, $k=1,...,n+1$ charged under a U(1) gauge field $A_{\mu}$  with Lagrangian density
\begin{equation}
{\cal L}_{\rm s}= \sum\limits_{k=1}^{n+1} \left[-D_{\mu} \Phi_k (D^{\mu}\Phi_k)^{*} - m^2 \Phi_k\Phi_k^* + \lambda_4 (\Phi_k\Phi_k^*)^2 - \lambda_6 (\Phi_k\Phi_k^*)^3 \right]\ \ , \ \
D_{\mu} = \partial_{\mu} - iqA_{\mu}
\end{equation}
in the 
background of $4$-dimensional spherically symmetric, static compact object that is a solution to the equations deriving from the following Lagrangian density
\begin{equation}
\tilde{{\cal L}}=\frac{{\cal R}}{16\pi G} + \beta^2 \Bigl(1 - \sqrt{1 + \frac{F_{\mu \nu}F^{\mu \nu}}{2 \beta^2}} \ \Bigr)  \  \ , \ \ F_{\mu\nu}=\partial_{\mu}A_{\nu} - \partial_{\nu}A_{\mu}  \ .
\end{equation}
$q$ is the gauge coupling, $m$ the mass of the scalar field and $\lambda_4 \geq 0$, $\lambda_6 \geq 0$ are the two self-interaction parameters of the scalar potential.
For $\beta\rightarrow \infty$ the Lagrangian density $\tilde{{\cal L}}$ is that of standard Maxwell's electrodynamics, while for $\beta < \infty$ the model describes Born-Infeld electrodynamics - in both cases coupled to standard Einstein gravity with Ricci scalar ${\cal R}$ and Newton's constant $G$.
\\\\
In the following, we will be interested in background solutions with U(1) gauge field
of the form
\be
\label{eq:ansatz_em}
               A_{\mu}{\rm d}x^{\mu} = V(r) {\rm d}t + Q_m \cos (\theta) {\rm d} \varphi
\ee
where $Q_m$ is a parameter that can be interpreted as the magnetic charge of the solution.
We also assume the space-time to be spherically symmetric and static~:
\be
\label{eq:ansatz_metric}
     {\rm d}s^2 = - N(r) {\rm d}t^2 + \frac{1}{N(r)} {\rm d}r^2 + r^2  \left({\rm d}\theta^2 + \sin^2\theta {\rm d}\varphi^2\right)\ \ , \ \ N(r)=1-\frac{2m(r)}{r}  \ .    
\ee
The background solution will then be given by a solution of the combined Einstein-gauge field equations resulting from the variation of $\tilde{\cal L}$ and using the Ans\"atze (\ref{eq:ansatz_em}) and (\ref{eq:ansatz_metric}). For the dynamical scalar field in this background solution, we will now distinguish two cases
\begin{itemize}
\item $Q_m=0$: In this case we can choose $n=0$ to obtain static and spherically symmetric solutions with scalar field Ansatz
\be
    \Phi_1 = \phi(r) e^{i \omega t} \ \ \ , \ \ \  \omega \in \mathbb{R} \ .
\ee
\item $Q_m\neq 0$: In this case we need $n=1$ to obtain static and spherically symmetric solutions with Ansatz \cite{Herdeiro:2024yqa}:
\be
    \Phi_1 = \phi(r)  \sin\left(\frac{\theta}{2}\right) e^{-i( \varphi/2 -\omega t)}\ \  \ , \ \ \  
		\Phi_2 = \phi(r)  \cos\left(\frac{\theta}{2}\right) e^{i( \varphi/2 +\omega t)} \ \ , \ \  \omega \in \mathbb{R} \ .
\ee
\end{itemize}
Note that the solutions fulfill the Dirac quantization condition~:
\be
                  q Q_m = \pm \frac{n}{2} \ \ , \ \ n = 0,1,2,\dots  \ .
\ee
The scalar field equation for $\phi(r)$ reads
\be
\label{eq:kg_dyon}
   \phi'' + \left(\frac{2}{r} + \frac{N'}{N} \right)\phi' + \frac{(\omega - qV)^2}{N^2} \phi - \frac{1}{2N} \frac{d U}{d \phi}
	= \frac{n}{2} \frac{\phi}{N r^2} \ \ ,
\ee
where the prime now and in the following denotes the derivative with respect to $r$ and $U(\phi)=m^2\phi^2 - \lambda_4\phi^4 + \lambda_6 \phi^6$ is the scalar field potential.
Asymptotically, the scalar field function has the following behaviour
\be
\label{eq:asymptotics_scalar}
\phi(r\rightarrow \infty)\sim \frac{\exp(-m_{\rm eff, \infty}^2 r)}{r} \ \ , \ \ m_{\rm eff, \infty}^2 = m^2-\Omega^2 
\ee
with $\Omega^2:=(\omega - q V_{\infty})^2$ where $V_{\infty}$ is the value of the electric potential at infinity: $V_{\infty}=V(r\rightarrow \infty)$. Hence, localized solutions to (\ref{eq:kg_dyon}) are only possible for $m^2 > (\omega - q V_{\infty})^2$. 
In the following, we will choose $m=1$, $\lambda_4=1$ and $\lambda_6=9/32$ unless otherwise stated. Note that one can think of the equation (\ref{eq:kg_dyon}) as describing a scalar 
field in an effective effective potential of the form
\begin{equation}
U_{\rm eff}=U  - \frac{\left(\omega-qV\right)^2}{N}\phi^2 + \frac{n}{2r^2}\phi^2 \ ,
\end{equation}
which is explicitly $r$-dependent for $n\neq 0$.

The physical parameters of the solutions that we are interested in are the Noether charge $Q_N$ and the average radius $\bar{R}$ of the solution. Within our Ansatz these read~:
\be
\label{eq:noether_radius}
Q_N=\int\limits_{r_0}^{\infty} {\rm d}r  \frac{2r^2 q V \phi^2}{N} \ \ \ \ \ \ , \ \  \ \ \ 
\bar{R}=\frac{\int\limits_{r_0}^{\infty} {\rm d}r r^2 (\phi(r))^3}{\int\limits_{r_0}^{\infty} {\rm d}r r^2 (\phi(r))^2} \ .
\ee
Moreover, the mass of the cloud is
\be
\label{eq:mass}
 M_Q = \int{\rm d}^3 x (T_j^j - T_0^0) = 
 \int \limits_{r_0}^{\infty} {\rm d} r r^2 \left(\frac{4q^2 V^2 \psi^2}{N} - 2U\right)     
\ee
In both (\ref{eq:noether_radius}) and (\ref{eq:mass}) $r_0=0$ for globally regular backgrounds and $r_0=r_h$ for black hole backgrounds.

\section{Scalar clouds on Reissner-Nordstr\"om black holes}
We will first investigate the case $\beta^2\rightarrow \infty$ for which only background black hole space-times are possible. This case has been discussed in detail in \cite{Brihaye:2020vce} for electrically charged Reissner-Nordstr\"om (RN) solutions and partly for electrically and magnetically charged RN solutions in \cite{Herdeiro:2024yqa}.
Here, we will present some additional results, and in particular give the domain of existence of scalarized electrically charged RN black holes in the $q$-$Q$-plane. 
\\\\
The general form of the RN solutions reads
\be
\label{reisner}
       N = 1 - \frac{2M}{r} + \frac{Q^2 + Q_m^2}{r^2}   \ \ , \ \ 
			 M = \frac{r_h}{2} \left(1 + \frac{Q^2 + Q_m^2}{r_h^2}\right) \ \ , \ \
\ee
where $Q$ is the electric charge and $Q_m$ the magnetic charge of the solution, respectively.
The associated electromagnetic field is
\be
			 V(r) = V_h + Q \left(\frac{1}{r_h} - \frac{1}{r}\right) \ \ ,
\ee
with $V_h=V(r_h)$ the value of the electric potential on the horizon with radius
$r=r_h=M+\sqrt{M^2 - Q^2- Q_m^2}$. For $M=r_h=\sqrt{Q^2 + Q_m^2}$, the solution is extremal
with vanishing Hawking temperature and near-horizon geometry AdS$_2\times S^2$. In the following, we will choose $V_h=0$ which results from the synchronization condition necessary to have scalar hair on charged black holes $\omega= qV_h$ \cite{Hong:2019mcj, Herdeiro:2020xmb} as well as our gauge choice $\omega=0$.

At the event horizon $r=r_h$ we want the scalar field to be regular and hence require the following boundary condition~:
\be
          N'(r_h) \phi'(r_h) = \frac{1}{2} \frac{dU}{d \phi}\left\vert_{r=r_h}  + \frac{n \phi(r_h)}{2 r_h^2}  \right.
\ee

Using $M = r_h/2+ (Q^2 + Q_m^2)/(2 r_h)$, a Taylor expansion of the scalar field in the variable $y\equiv r-r_h$ gives
\be
     \phi = \phi_h + \phi_1 y 
		               + \phi_2 y^2  + {\cal O}(y^3)  \ ,
\ee 
where
\be
\phi_1 = \frac{r_h^3}{r_h^2 - (Q^2+Q_m^2)} \left(\frac{\dot{U}(r_h)}{2} + \frac{n \phi_h}{2 r_h^2} \right) \ ,
\ee
and 
\be
   \phi_2 = \frac{n^2 \phi_0 r_h^2}{16(Q^2 + Q_m^2 - r_h^2)^2} 
	+ \frac{n r_h^2(r_h^2(\dot U + \phi_0 \ddot U)-4)}{16(Q^2 + Q_m^2 - r_h^2)^2}
	+ \frac{r_h^2
	\Bigl(\dot U(\phi_h) (r_h^4 \ddot U(\phi_h) - 4 (Q^2 + Q_m^2)) - 4 \phi_h q^2 Q^2 \Bigr)} {16(Q^2+ Q_m^2-r_h^2)^2}   \ ,
\ee
where $\phi_h=\phi(r_h)$ and the dot denotes the derivative of the potential with respect to $\phi$. 
Inserting this expansion into 
(\ref{eq:kg_dyon}) and linearizing leads to the observation that the scalar field 
has effective mass
\be
         m_{\rm eff}^2 = m^2 - \frac{q^2 Q^2}{r_h^2} \ \ 
\ee
close to the horizon of the RN black hole. We need to require $m_{\rm eff}^2 \geq 0$ for well behaved solutions. In other words, we need to make sure that the electric potential related to the black hole does not become so large that it becomes possible to create particles of mass $m$ close to the horizon.

Hence, we find two constraints on the parameters in our model:

\begin{enumerate}
    \item  $Q^2 + \left(\frac{n}{2 q}\right)^2 < r_h^2$ which results from the requirement of considering black holes as opposed to naked singularities,
    \item $0 < Q < \frac{m r_h}{q}$ which is necessary to have a real effective mass close to the horizon (see discussion above).
\end{enumerate}

\begin{figure}[h]
\begin{center}
\includegraphics[width=8cm]{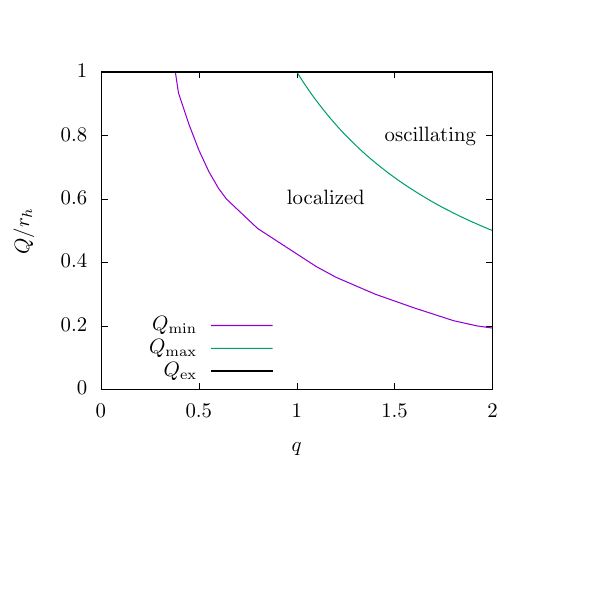}
\hfill
\includegraphics[width=8cm]{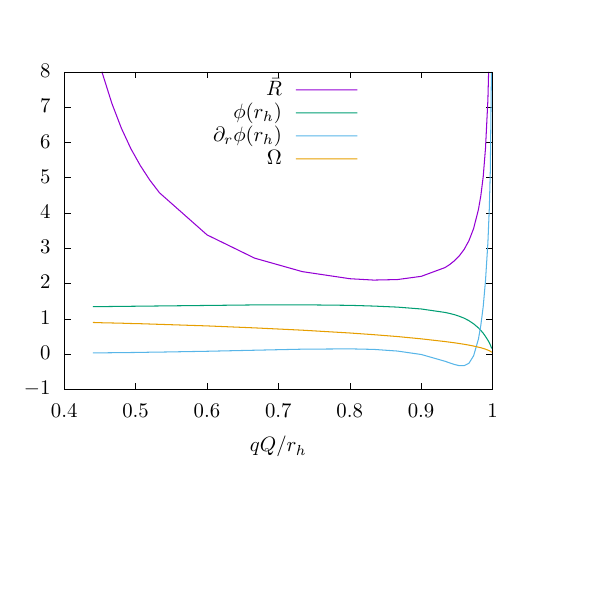}
\vspace{-2cm}
\caption{The domain of existence of scalarized electrically charged RN black holes in the $q$-$Q$-plane for $m=1$ and $r_h=0.15$. $Q_{\rm min}$ (purple) corresponds to the minimal value of the black hole charge, $Q_{\rm max}=mr_h/q$ (green) to the maximal possible charge beyond which localized scalar hair exists, while $Q_{\rm ex}=r_h$ (black) corresponds to the extremal RN solution. For values of $Q_{\rm max} < Q < Q_{\rm ex}$ black holes with spatially oscillating scalar field exist (left). The mean radius $\bar{R}$ (purple), the value of the scalar field on the horizon $\phi(r_h)$ (green), the value of the derivative of the scalar field on the horizon $\partial_r\phi(r_h)=\phi'(r_h)$ (blue) as well as $\Omega$ (orange) in function of $qQ/r_h$ (right).
\label{fig:domainRN}
}
\end{center}
\end{figure}

\subsection{Scalar clouds on electrically charged RN black holes}
\label{subsection:RNq}
\begin{figure}[h]
\begin{center}
\includegraphics[width=8cm]{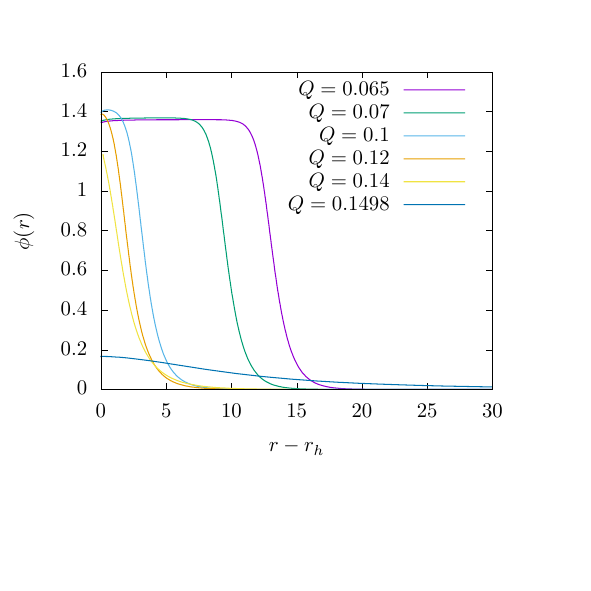}
\hfill
\includegraphics[width=8cm]{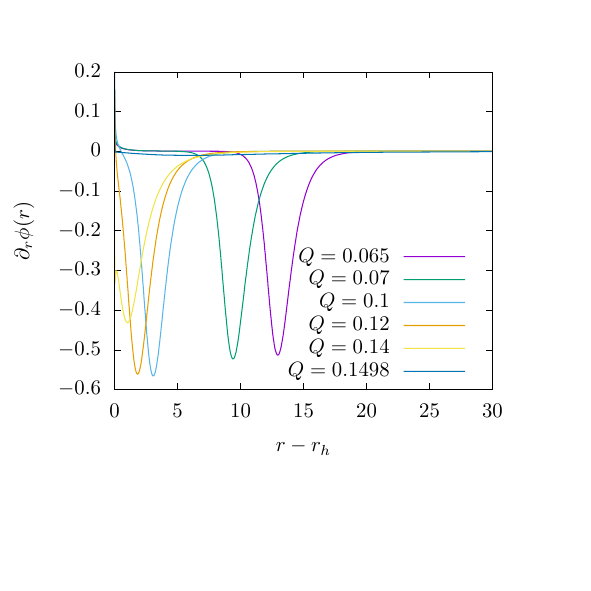}
\vspace{-2cm}
\caption{The profiles of $\phi$ (left) and of $\partial_r \phi=\phi'$ (right) for $q=1$, $r_h = 0.15$ 
and different values of $Q$.
\label{fig:phi_phip_Q}
}
\end{center}
\end{figure}

In the following, we have chosen $m=1$, $r_h=0.15$ and $Q_m=0$, i.e. we will study scalar clouds on electrically charge RN black holes with $Q\neq 0$. The parameters determining the existence of solutions are then the gauge coupling $q$, which we assume to be positive in the following, and the charge of the RN black hole $Q$. The domain of existence of scalarized RN black holes in the $q$-$Q$-plane is limited by the constraints mentioned above. We have determined this domain by constructing scalarized RN black holes numerically and our results are shown in Fig.\ref{fig:domainRN} (left). For $q \lesssim 0.39$ we find that no scalarized black holes exist at all independent of the value of $Q$. The constraint tells us that for $n=0$ and $Q$, $r_h$ fixed, there is no lower bound on the parameter $q$. Hence, a non-linear phenomenon is at play here that we will explain below. For $q \gtrsim 0.39$ we find that scalarized solutions exist on a finite interval of the black hole charge $Q$. For $Q\rightarrow Q_{\rm min}$ we find that the scalar field spreads over the whole interval of $r$ and is no longer localized on the horizon.
    This can be seen in Fig.\ref{fig:domainRN} (right), where we give the mean radius $\bar{R}$, the value of the scalar field on the horizon $\phi(r_h)$, the value of the derivative of the scalar field on the horizon $\phi'(r_h)$ as well as $\Omega$ in function of $qQ/r_h$ for $Q=0.09$ and $r_h=0.15$. Clearly at sufficiently small $q$, the mean radius diverges, while $\Omega$ and $\phi(r_h)$ stay perfectly finite. This can also be seen in Fig.\ref{fig:phi_phip_Q}, where we show the profiles of $\phi(r)$ and $\phi'(r)$ for $q=1$ and $r_h=0.15$. Decreasing $q$ we find that the scalar field starts to form a {\it thin wall} at some intermediate value of the radial coordinate $r=r_0$ such that the scalar field $\phi(r)$ is approximately constant and non-vanishing on the interval $r\in [0:r_0]$ and $\phi(r)\equiv 0$ for $r > r_0$. The value of the scalar field function inside the wall is the value of the {\it false vacuum} of the scalar field as discussed previously in \cite{Brihaye:2020vce}. Increasing the value of $Q$ for a fixed $q$ we find that for $1 \geq q \gtrsim 0.39$ localized scalar fields exist for $Q\in [Q_{\rm min}:Q_{\rm ex}]$ where $Q_{\rm ex}=r_h$ is the maximal possible charge for a RN black hole, i.e. the extremal limit. For $q > 1$ localized solutions exist for $Q\in [Q_{\rm min}:Q_{\rm max}$], where $Q_{\rm max}=mr_h/q$ is determined by the requirement of real effective mass of the scalar field. For $q > 1$ and $Q \in [Q_{\rm max}:Q_{\rm ex}]$ we find solutions that oscillate spatially outside the horizon which is not surprising as $m_{\rm eff}^2 < 0$.

\subsection{Scalar clouds on electrically and magnetically charged RN black holes}
\begin{figure}[h]
\begin{center}
\includegraphics[width=8cm]{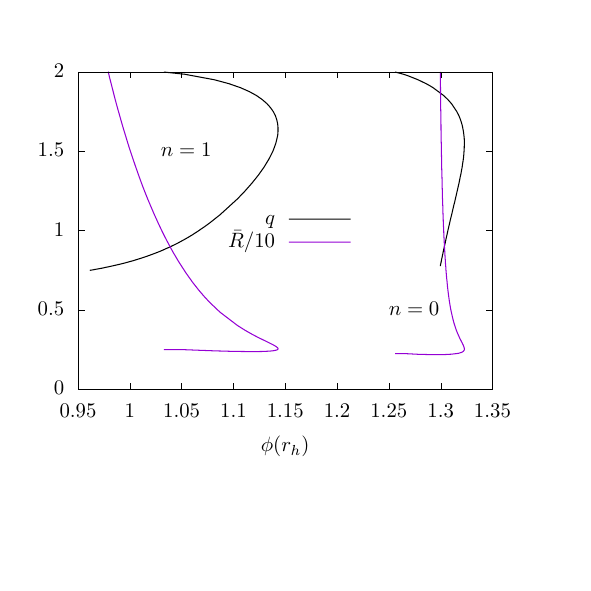}
\hfill
\includegraphics[width=8cm]{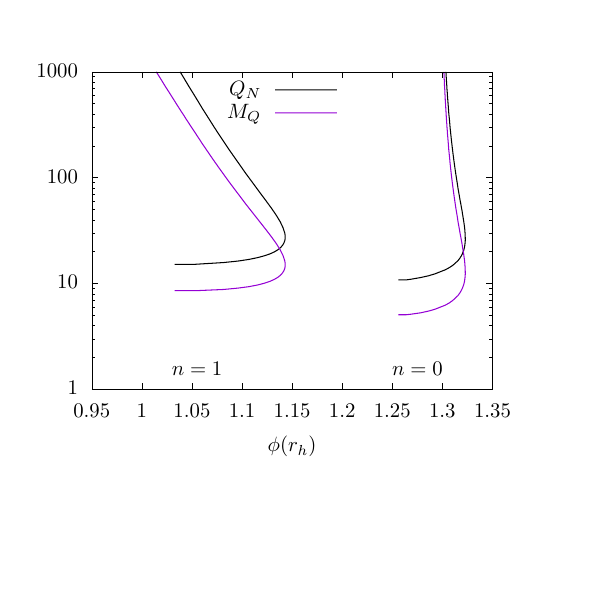}
\vspace{-2cm}
\caption{{\it Left}: The value of the gauge coupling $q$ (black) and the mean radius $\bar{R}$ (purple) in function of $\phi(r_h)$  for scalar clouds on doubly charged RN solutions ($n=1$) with
$Q=0.5$ and $r_h=1$. We also show the corresponding values for scalar clouds on electrically charged RN solutions ($n=0$).
{\it Right}: The value of the Noether charge $Q_N$ (black) and the mass $M_Q$ (purple) in function of $\phi(r_h)$  for scalar clouds on doubly charged RN solutions ($n=1$) with
$Q=0.5$ and $r_h=1$. We also show the corresponding values for scalar clouds on electrically charged RN solutions ($n=0$).
\label{fig:data_dyon}
}
\end{center}
\end{figure}

\begin{figure}[h]
\begin{center}
\includegraphics[width=8cm]{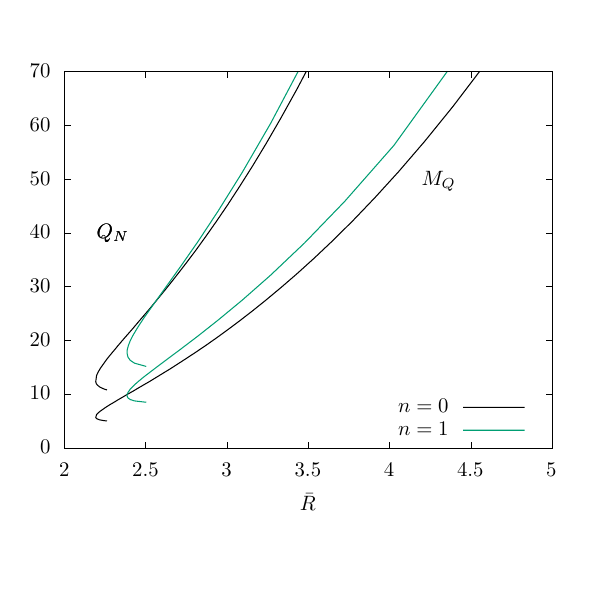}
\vspace{-1cm}
\caption{We show the mass $M_Q$ and the Noether charge $Q_N$ of a scalar cloud on
a charged RN black holes with $r_h=1$ and $Q=0.5$ in function of the mean radius $\bar{R}$. The RN black holes are electrically charged ($n=0$, black) or doubly, i.e. electrically and magnetically charged ($n=1$, green).
\label{fig:R_Q_M_dyon}
}
\end{center}
\end{figure}

We will now discuss the influence of the additional magnetic charge on the results given above, i.e. we now choose $Q\neq 0$ and $Q_m=\pm n/(2q)\neq 0$.
Several parameters characterizing the solutions for $n=0$ and $n=1$ are given in Fig. \ref{fig:data_dyon}, where we show the value of $q$, the mean radius $\bar{R}$, the mass $M_Q$ and the Noether charge $Q_N$  in function of the value of the scalar field on the horizon, $\phi(r_h)$, for $r_h=1$ and $Q=0.5$. We find that scalar clouds on doubly charged RN solutions require the scalar field to be smaller on the horizon as compared to the electrically charged case. Apart from that we find qualitatively similar dependence on $\phi(r_h)$. For both the $n=0$ as well as the $n=1$ solutions two branches of solutions in $\phi(r_h)$ exist. This had been reported before for $n=0$ \cite{Brihaye:2020vce}. The first branch has higher values of $q$ and ends at $q=2$
due to the requirement $Q < mr_h/q$ with $m=1$, $r_h=1$ and $Q=0.5$. This branch corresponds to lower $\bar{R}$, $Q_N$ and $M_Q$, respectively. On this first branch, the increase of $\phi(r_h)$ leads to a slight increase in the mean radius $\bar{R}$, the mass $M_Q$ and the Noether charge $Q_N$. Having reached the maximal possible value of $\phi(r_h)$, we find a second branch of solutions on which $\bar{R}$, $M_Q$ and $Q_N$ increase further when decreasing $\phi(r_h)$. At the same time $q$ decreases. 
This branch stops at a finite critical value of $\phi(r_h)$ and is connected to the phenomenon of the formation of a thin wall described in \ref{subsection:RNq}. The presence of the magnetic charge does not change this qualitative phenomenon. 
In Fig.\ref{fig:R_Q_M_dyon} we show the mass $M_Q$ and the Noether charge $Q_N$ in function of the mean radius $\bar{R}$. On the first branch of solutions 
the mass $M_Q$ and Noether charge $Q_N$ decrease with increasing mean radius, while on the second branch of solutions $M_Q$ and $Q_N$ increase with the increase of $\bar{R}$.
We observe that $n=0$ solutions can have slightly smaller
mean radius than the $n=1$ solution with minimal possible $\bar{R}$ at the bifurcation of the two branches of solutions. Interestingly, we find that there exists a small interval in $\bar{R}$ close to $\bar{R}\approx 2.5$ for which the $n=1$ solutions have smaller values of $M_Q$ than the corresponding $n=0$ solutions for the same value of $\bar{R}$. For $\bar{R} \gtrsim 2.5$ the mass of the $n=1$ solutions is always higher than that of the $n=0$ solution. Moreover, the $n=1$ solutions have smaller
Noether charge $Q_N$ on an interval of $\bar{R}$ between $\approx 2.5$ and $\approx 2.7$. For $\bar{R}\gtrsim 2.7$ the Noether charge of the $n=1$ solution is always larger in value than that of the $n=0$ solution.

\section{Scalarized Born-Infeld black holes and stars}
In the following, we will only consider solutions that carry electric charge, i.e. we choose $Q_m=0$. For $\beta^2$ finite, the solutions of the gravity-gauge field equations then fulfill 
\be
\label{eq:BI}
    m' = \beta\left(\sqrt{Q^2 + \beta^2 r^4} - \beta r^2\right) \ \ , \ \ V' = \frac{\beta Q}{\sqrt{Q^2 + \beta^2 r^4}} \ .
\ee
There are black holes as well as globally regular, soliton-like solutions to these equations. 

\subsection{Scalarized Born-Infeld black holes}
Born-Infeld black holes (BIBHs) are solutions to the equations (\ref{eq:BI}) with the boundary condition  $m(r_h) = r_h/2$. Note that these solutions have an extremal limit with $N(r_h)=N'(r_h)=0$ for $Q^2=r_h^2 + 1/(4\beta^2)$. 
Close to the horizon at $r=r_h$ we find~:
\be
N(r) = \frac{r-r_h}{r_h} \left(1 + 2 \beta^2 r_h^2 - 2 \beta \sqrt{Q^2 + \beta^2 r_h^4}\right) + {\cal O}(r-r_h)^2 \ ,
\ee
and
\be
      V(r) = V(r_h) + \frac{\beta Q}{\sqrt{Q^2 + \beta^2 r_h^4}}(r-r_h)  + {\cal O}(r-r_h)^2 \ .
\ee
The resonance condition $\omega=qV_h$ is the same as in the RN case - the scalar field does not backreact. We have constructed BIBHs for several choices of $\beta$, $Q$, $q$ and $r_h$. We do not find a huge qualitative difference and/or new phenomena appearing in this case, at least not for the values of coupling constants that we have studied. Especially, we note that $\beta$ has to be chosen very small to see any effect at all. Hence, we will not discuss these solutions in more detail here, but will concentrate on the solutions that do not exist in the $\beta=\infty$ limit: globally regular solutions.

\subsection{Scalarized Born-Infeld stars}
Born-Infeld stars (BIS) are solutions to (\ref{eq:BI}) with the boundary condition $m(0)=0$. Close to $r=0$ the metric function and the electric potential have the following form
\be
N(r) = 1 - 2\beta Q + {\cal O}(r^2) \ \ \ , \ \ \ V(r) = V(0) + \beta r - \frac{\beta^3}{10 Q^2} r^5 + {\cal O}(r^9) \ .
\ee
Note that in order to have globally regular solutions we need to require that
$\beta < 1/(2Q)$. Moreover, the solution has a non-vanishing electric field $E_r \propto \beta$ at the origin $r=0$. If we now want this BIS to carry a non-trivial scalar field, we require the following conditions~:
\be
\label{eq:BCBIS}
    \phi(0) = \phi_0 \ \ , \ \ \phi'(0) = 0 \ \ , \ \ \phi(r \to \infty) = 0 \ \ .
\ee
Note that these are three conditions on $\phi(r)$ which means that $V(0)$ (the only other, a priori, free parameter in the system) is fixed by the conditions (\ref{eq:BCBIS}) and cannot be chosen freely.
\\

\subsubsection{Numerical results}
\begin{figure}[h]
\begin{center}
\includegraphics[width=8cm]{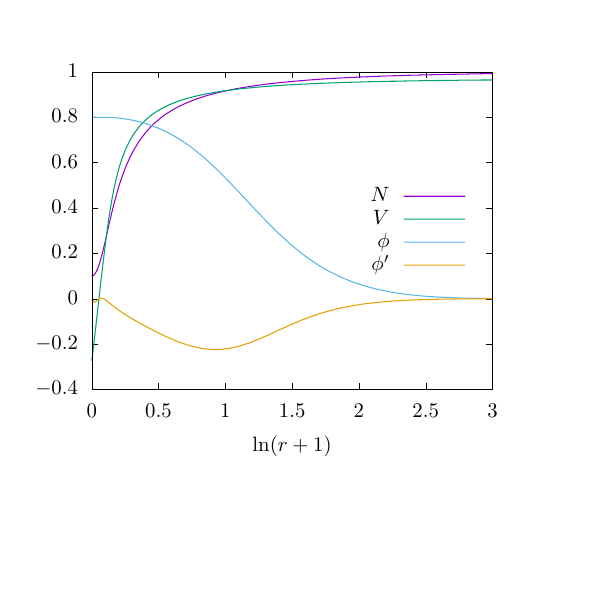}
\hfill
\includegraphics[width=8cm]{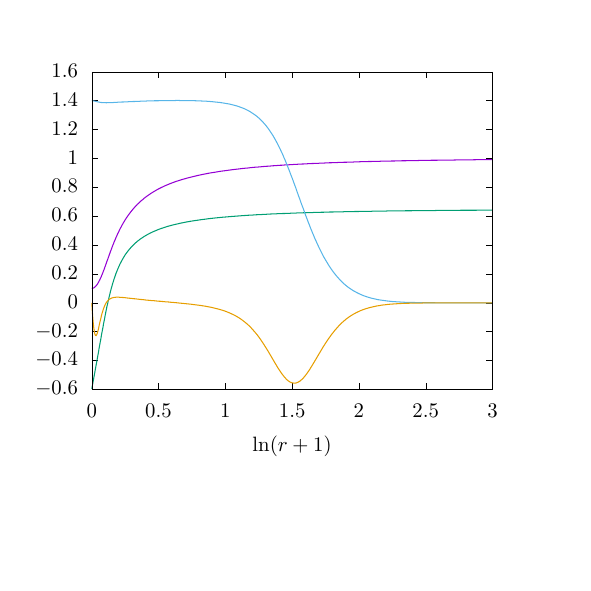}
\vspace{-2cm}
\caption{We show the profiles of the scalar field function $\phi(r)$, the derivative of the scalar field function $\phi'(r)$, the metric function $N(r)$ as well as the electric potential $V(r)$ of a typical scalarized BI star. Here $\beta=5, Q=0.09$, $q=1.0$
and $\phi(0) = 0.8$ (left, first branch of solutions) and $\phi(0)=1.4$ (right, second branch of solutions).
\label{fig:profiles_BIsoliton}
}
\end{center}
\end{figure}

The profiles of the scalar field $\phi(r)$, the scalar field derivative $\phi'(r)$, the metric function $N(r)$ and the electric potential $V(r)$ of typical scalarized BI stars with $\beta=5$, $Q=0.09$ and $q=1.0$  is shown in Fig. \ref{fig:profiles_BIsoliton}. This choice of $\beta Q=0.45$ is close to the maximal possible limit of $\beta Q =0.5$ and implies $N(0)=0.1$. The profiles on the left of  Fig. \ref{fig:profiles_BIsoliton} are for $\phi(0)=0.8$ and belong to the first branch of solutions (see more details below), while the profiles on the right of Fig. \ref{fig:profiles_BIsoliton}
are for $\phi(0)=1.4$ and belong to the second branch of solutions. The metric function $N(r)$ does not change, however the electric potential $V(r)$ shifts quantitatively. Since only potential differences are physically meaningful, the qualitative
profile, of course, does not change and the potential difference between the origin and infinity is the same.

\begin{figure}[h]
\begin{center}
\includegraphics[width=8cm]{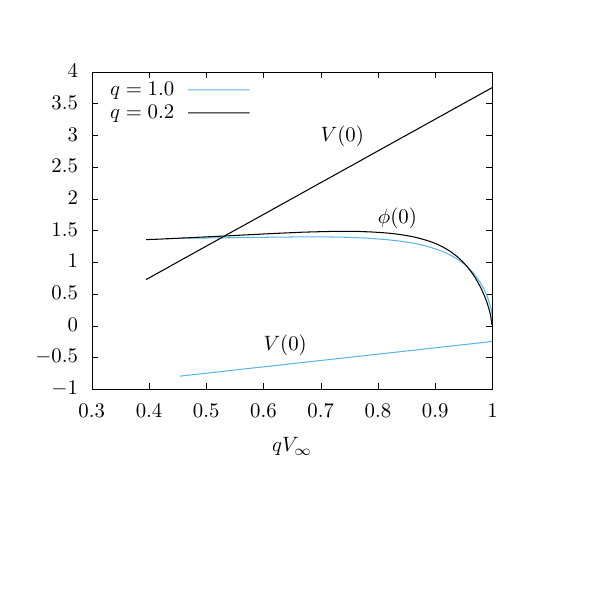}
\hfill
\includegraphics[width=8cm]{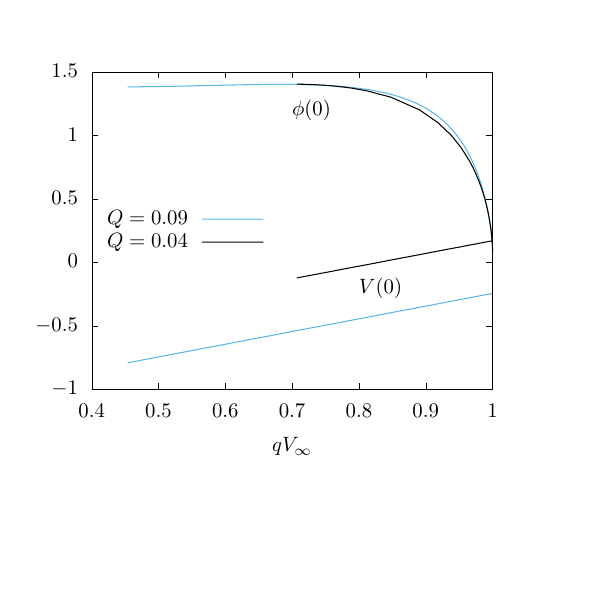}
\vspace{-2cm}
\caption{{\it Left}: We show the value of the scalar field function and the electric potential at the origin, $\phi(0)$ and $V(0)$, respectively, for $Q=0.09$ and $q=1.0$ (blue) and $q=0.2$ (black) in function of $qV_{\infty}$.
{\it Right}: Same as left, but for $q=1.0$ and two different values of $Q$: $Q=0.09$ (blue) 
and $Q=0.04$ (black).
\label{fig:BIsoliton_Q009}
}
\end{center}
\end{figure}

\begin{figure}[h]
\begin{center}
\includegraphics[width=8cm]{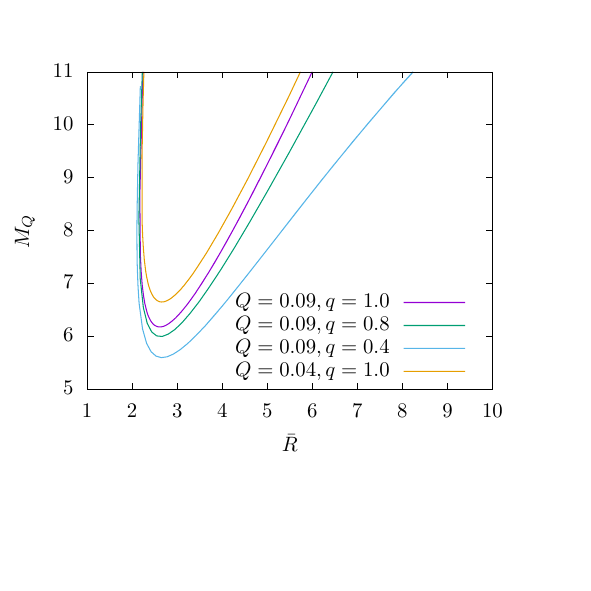}
\hfill
\includegraphics[width=8cm]{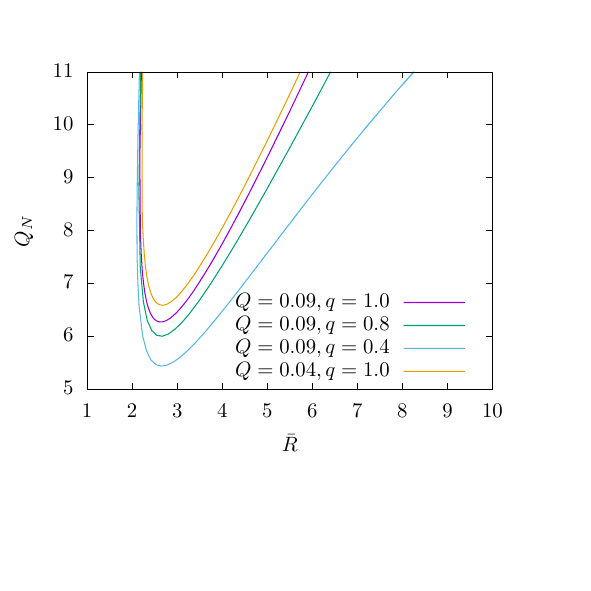}
\vspace{-2cm}
\caption{{\it Left}: We show the value of the mass $M_Q$ as function of the mean radius $\bar{R}$ for several values of $Q$ and $q$.
{\it Right}: We show the value of the Noether charge $Q_N$ as function of the mean radius $\bar{R}$ for several values of $Q$ and $q$.
\label{fig:BI_R_QN_MQ}
}
\end{center}
\end{figure}

To demonstrate how $V(0)$ changes, we show in Fig. \ref{fig:BIsoliton_Q009} the values of $\phi(0)$ and $V(0)$ in dependence on $qV_{\infty}$. As $qV_{\infty}$ increases, so does $V(0)$ and we seem to find a nearly linear increase.
At the same time, $\phi(0)$ decreases. This is true for both $q=0.2$ and $q=1.0$ with BI star charge $Q=0.09$ (see left) as well
as for $q=1.0$ and $Q=0.04$ (see right). Interestingly, the slope of the $qV_{\infty}$-$V(0)$-curve increases when lowering $q$ at fixed $Q$, however stays more or less the same when keeping $q$ fixed and changing $Q$. All curves end at $qV_{\infty}=1$, while the minimal possible value of $qV_{\infty}$
has to be determined numerically and is a non-linear phenomenon.
We find that the minimal value of $qV_{\infty}$ seems to depend strongly on the choice of $Q$ - decreasing strongly when increasing $Q$ - while a change in $q$ only leads to a small change in this value. The existence of this minimal value of $qV_{\infty}$ is related to the formation of a thin wall as can be seen in Fig.\ref{fig:profiles_BIsoliton} (right). This is very similar to our results in the black hole case. The formation of this wall is related to the existence of a second branch of solutions in $\bar{R}$, see Fig \ref{fig:BI_R_QN_MQ}. In this figure, we show the mass $M_Q$ (left) and the Noether charge $Q_N$ (right), respectively, in function of the mean radius $\bar{R}$ for several choices of $Q$ and $q$. We find that two branches of solutions exist and that these meet at a minimal value of $M_Q$ and $Q_N$, respectively, and at a value of $\bar{R}$ close to the minimal possible value. On the first branch of solutions, increasing $\phi(0)$ from $\phi(0)=0$ we find that $M_Q$ and $Q_N$ decrease with the decrease of $\bar{R}$. Increasing $\phi(0)$ further leads to a sharp increase in $M_Q$ and $Q_N$, while $\bar{R}$ decreases only slightly to the minimal possible value of the mean radius. While we see
a quantitative dependence on the values of $Q$ and $q$ on the first branch of solutions, this is different on the second branch, where $M_Q$ and $Q_N$ increase strongly close to $\bar{R}\approx 2$ and we see only slight dependence on $q$ and $Q$.
On the first branch, we find that increasing $q$ for a fixed $Q$ increases both $M_q$ and $Q_N$, while fixing $q$ and increasing $Q$ has the opposite effect. We conclude that choosing $Q$ small and $q$ large leads to the most compact scalar clouds.

\section{Conclusions}
In this paper, we have reported results on the numerical construction of charged stars and black holes that carry non-trivial scalar hair. The scalar clouds surrounding these compact objects can be thought off as being made of a number $Q_N$ of scalar bosons. The coupling to a U(1) gauge field leads to these scalar bosons being charged and interacting with the charge of the compact object. Hence, the mean radius and mass $M_Q$ of the cloud depend crucially on the choice of the charge of the compact object $\sqrt{Q^2 + Q_m^2}$ as well as on the gauge coupling $q$. 
We find that in general two branches of solutions in the mean radius $\bar{R}$ exist.
On the first branch the mass $M_Q$ and Noether charge $Q_N$ decrease with decreasing $\bar{R}$ for both the black hole as well as BI star solutions. The qualitative behaviour on the second branch is different for black holes as compared to BI stars:
while the mass and Noether charge decrease further and the radius increases for black holes, we see a sharp increase in the mass and Noether charge with slightly decreasing radius for BI stars. In both cases, we see the formation of a thin wall
in the scalar field profile that separates the false vaccum interior from the true vaccum exterior. 
Obviously, it is interesting to see whether the backreaction of the scalar field on the BI star could lead to interesting new phenomena such as the formation of "wavy scalar hair" that was observed previously for black holes \cite{Brihaye:2021phs} and boson stars \cite{Brihaye:2021mqk}. 

Finally let us not that the scalarized BI star is an interesting new globally regular solution that carries charge and a finite electric field at the origin. 
That the gravitational field of such a solution has interesting effects is shown in the Appendix where we discuss the change of a magnetic dipole in the space-time of such a star. 
\\\\

\clearpage

\clearpage

\section*{Appendix: Magnetic multipoles in the space-time of a scalarized BI star}

\begin{figure}[htb]
\begin{center}
\includegraphics[width=8cm]{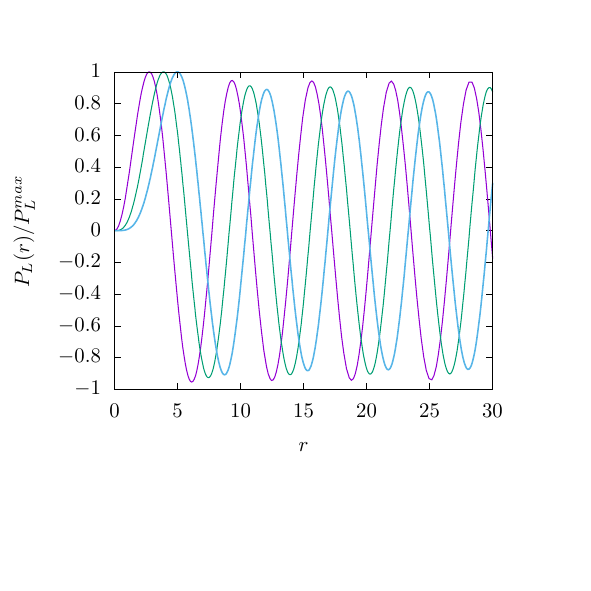}
\hfill
\includegraphics[width=8cm]{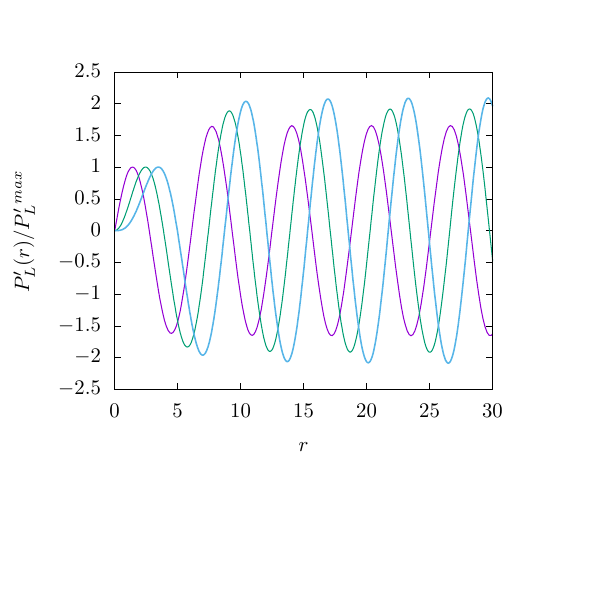}
\end{center}
\vspace{-1cm}
\caption{{\it Left}: We show the function $P_L(r)$ (see (\ref{eq:MMP})) for $L=1$ (purple), $L=2$ (green) and $L=3$ (blue). {\it Right}: We show the function $P'_L(r)$ for $L=1$ (purple), $L=2$ (green) and $L=3$ (blue). Note that in both figures, the function is divided by the value of the first maximum $P_L^{max}$ and $P^{\prime max}_L$, respectively.
\label{fig:MM_PPp}
}
\end{figure}

Maxwell's equations do not possess regular solutions for magnetic multipoles in an asymptotically flat black hole space-time background as the solutions either diverge at $r\to \infty$ or are singular for $r\rightarrow r_h$. (See \cite{Herdeiro:2016xnp} for solutions in asymptotically AdS).
Recently, a model for boson stars in Anti-de Sitter has been suggested with Ansatz for the electromagnetic field of the form \cite{Herdeiro:2024myz}~:
\be
            A_{\mu} {\rm d}x^{\mu} = P(r , \theta) e^{i \omega t} {\rm d}  \varphi \ \ .
\ee
Using this purely magnetic Ansatz, 
the relevant Maxwell equation can be separated using $P(r, \theta) = P(r) Z(\theta)$ which gives the two following equations:
\be
\label{eq: magnetic}
             (N P')' +  \left(\frac{\omega^2}{N}   +  \frac{1}{r^2}  \tilde K(\theta)\right)P = 0 \ \ , \ \
							 \tilde K(\theta) =\left(\frac{\sin \theta}{Z(\theta)}\right) \left(\frac{Z'}{\sin \theta}\right)' \ ,
\ee\
where the prime in the equation for $P$ denotes the derivative with respect to $r$, while in the equation for $Z$ it denotes the derivative with respect to $\theta$.
$N(r)$ is the metric function associated to the spherically symmetric space-time and $\tilde{K}(\theta)$ is a $\theta$-dependent factor. 
Choosing 
\be
\label{eq:angular_MM}
        Z(\theta) = \sin(\theta) \frac{d {\cal P}_L(\cos(\theta)}{d \theta} \,
\ee
where ${\cal P}_L$ is the Legendre polynomial of degree $L$, $\tilde{K}=-L(L+1)$ is constant. 
For example, for $L=1$, i.e. $\tilde{K}=-2$, equation (\ref{eq:angular_MM}) is solved by $Z = (\sin \theta)^2$. 
Note also that, in fact, a rescaling of the radial coordinate allows to set  $\omega = 1$ without loosing generality. 
The equation for $P$ can be solved in terms of Bessel functions for $N(r)\equiv 1$, i.e. in Minkowski space-time:
\be
\label{eq:MMP}
               P_L(r) = \sqrt r \  J_{L+1/2}(r)
\ee
$P_L(r)$ and $P_L'(r)$ are shown in Fig. \ref{fig:MM_PPp} for $L=1,2,3$.

\begin{figure}[h]
\begin{center}
\includegraphics[width=10cm]{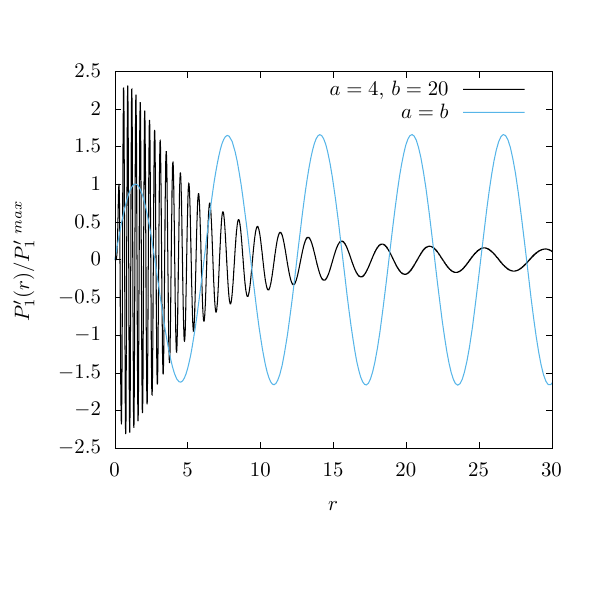}
\vspace{-1cm}
\caption{We show the magnetic dipole field $P'_1(r)$ in the background of flat space-time ($a=b$; blue) and in the background of a scalarized BI star ($a=4,b=20$; black). $P'_1(r)$ is normalized by the value of the first maximum $P^{\prime max}_1$.
\label{fig:field_mag_alpha}
}
\end{center}
\end{figure}

In the following, we want to demonstrate how these functions change when the background is that of a scalarized BI star instead of flat space-time. 
For simplicity we choose
\be 
\label{eq:n_para}
N(r) = (a^2 + r^2)/(b^2+r^2) \ \ ,  \ \ a,b  \ \  {\rm constants}
\ee
which is a very good fit for the metric function of such a star. We have then solved the $P$ equation (\ref{eq: magnetic}) using (\ref{eq:n_para}).
Our results for the magnetic dipole field $P'_1$ is given Fig.\ref{fig:field_mag_alpha}.
We show the solution in the flat space-time background ($a=b$) as well as in the background
of the scalarized BI star with $a=4$, $b=20$. The BI star increases the frequency and amplitude strongly close to the star with both the former and the latter decreasing with distance from the star. 

\begin{thebibliography}{99}


\bibitem{RN} H. Reissner: \"Uber die Eigengravitation des elektrischen Feldes nach der Einsteinschen Theorie, Annalen der Physik \textbf{355} (1916), 106; G. Nordstr\"om, On the Energy of the Gravitational Field in
Einstein’s Theory, Verhandl. Koninkl. Ned. Akad. Wetenschap., Afdel. Natuurk., Amsterdam. \textbf{26}
(1918), 1201.
\bibitem{jebsen} J. T. Jebsen:\"Uber die allgemeinen kugelsymmetrischen L\"osungen der Einsteinschen 
Gravitationsgleichungen im Vakuum, Arkiv f\"or Matematik, Astronomi och Fysik \textbf{15} (1921), 1; G. D. Birkhoff, in
Relativity and Modern Physics, Harvard University Press, Cambridge, Massachusetts (1923).
\bibitem{israel} W. Israel, Event horizons in static vacuum space-times, Phys. Rev. \textbf{164} (1967), 1776.



\bibitem{carter1971} B. Carter, Axisymmetric Black Hole Has Only Two Degrees of Freedom, Phys. Rev. Lett. \textbf{26} (1971), 331; The vacuum black hole uniqueness theorem and its conceivable generalisations, Proceedings
of the 1st Marcel Grossmann meeting on general relativity (1977), 243.
\bibitem{robinson} D. Robinson, Uniqueness of the Kerr Black Hole, Phys. Rev. Lett. \textbf{34} (1975), 905.
\bibitem{heusler} M. Heusler, Stationary Black Holes: Uniqueness and Beyond, Liv. Rev. Rel. \textbf{1} (1998).
\bibitem{MTW} C. Misner, K. Thorne, and J. Wheeler, Gravitation, W. H. Freeman and Company, (1973).

\bibitem{Born:1934gh}
M.~Born and L.~Infeld: Foundations of the new field theory,
Proc. Roy. Soc. Lond. A \textbf{144} (1934) no.852, 425.


\bibitem{Ayon-Beato:1998hmi}
E.~Ayon-Beato and A.~Garcia: Regular black hole in general relativity coupled to nonlinear electrodynamics,
Phys. Rev. Lett. \textbf{80} (1998), 5056.

\bibitem{Ayon-Beato:1999kuh}
E.~Ayon-Beato and A.~Garcia: New regular black hole solution from nonlinear electrodynamics,
Phys. Lett. B \textbf{464} (1999), 25.


\bibitem{Ayon-Beato:1999qin}
E.~Ayon-Beato and A.~Garcia: Nonsingular charged black hole solution for nonlinear source,''
Gen. Rel. Grav. \textbf{31} (1999), 629.

\bibitem{Bronnikov:2000vy}
K.~A.~Bronnikov: Regular magnetic black holes and monopoles from nonlinear electrodynamics,
Phys. Rev. D \textbf{63} (2001), 044005.

\bibitem{Burinskii:2002pz}
A.~Burinskii and S.~R.~Hildebrandt: New type of regular black holes and particle - like solutions from NED,
Phys. Rev. D \textbf{65} (2002), 104017.

\bibitem{Yang:2022qoc}
Y.~Yang: Dyonically charged black holes arising in generalized Born\textendash{}Infeld theory of electromagnetism,
Annals Phys. \textbf{443} (2022), 168996.

\bibitem{Dymnikova:2004zc}
I.~Dymnikova: Regular electrically charged structures in nonlinear electrodynamics coupled to general relativity,from nonlinear electrodynamics, Phys. Rev. D 65, 104017 (2002


\bibitem{Lobo:2006xt}
F.~S.~N.~Lobo and A.~V.~B.~Arellano: Gravastars supported by nonlinear electrodynamics,
Class. Quant. Grav. \textbf{24} (2007), 1069.



\bibitem{Herdeiro:2015waa}
C.~A.~R.~Herdeiro and E.~Radu: Asymptotically flat black holes with scalar hair: a review,
Int. J. Mod. Phys. D \textbf{24} (2015) no.09, 1542014

\bibitem{Hong:2019mcj}
J.~P.~Hong, M.~Suzuki and M.~Yamada: Charged black holes in non-linear Q-clouds with O(3) symmetry,
Phys. Lett. B \textbf{803} (2020), 135324.

\bibitem{Herdeiro:2020xmb}
C.~A.~R.~Herdeiro and E.~Radu: Spherical electro-vacuum black holes with resonant, scalar $Q$-hair,
Eur. Phys. J. C \textbf{80} (2020) no.5, 390.

\bibitem{Brihaye:2020vce}
Y.~Brihaye and B.~Hartmann: Strong gravity effects of charged Q-clouds and inflating black holes,
Class. Quant. Grav. \textbf{38} (2021) no.6, 06LT01.




\bibitem{Herdeiro:2024yqa}
C.~Herdeiro, E.~Radu and Y.~Shnir: Reissner-Nordstr\"om dyonic black holes with gauged scalar hair,
[arXiv:2406.10643 [hep-th]].



\bibitem{Herdeiro:2024myz}
C.~Herdeiro, H.~Huang, J.~Kunz and E.~Radu: Einstein-(complex)-Maxwell static boson stars in AdS,
[arXiv:2405.10671 [gr-qc]].

\bibitem{Brihaye:2021phs}
Y.~Brihaye and B.~Hartmann: Spherically symmetric charged black holes with wavy scalar hair,
Class. Quant. Grav. \textbf{39} (2022) no.1, 015010

\bibitem{Brihaye:2021mqk}
Y.~Brihaye and B.~Hartmann: Boson stars and black holes with wavy scalar hair,
Phys. Rev. D \textbf{105} (2022) no.10, 104063


\bibitem{Herdeiro:2016xnp}
C.~Herdeiro and E.~Radu: Einstein\textendash{}Maxwell\textendash{}Anti-de-Sitter spinning solitons,
Phys. Lett. B \textbf{757} (2016), 268-274
	
	
\end{thebibliography}
\end{document}